# The Proximity Effect on the Lyman $\alpha$ Forest Due to a Foreground QSO

A. Fernández-Soto[1], X. Barcons[2], R. Carballo[1] and J.K. Webb[3]

[1] Departamento de Física Moderna. Universidad de Cantabria. Avenida de Los Castros s/n. E-39005 Santander, Spain
[2] Instituto de Física. Consejo Superior de Investigaciones Científicas - Universidad de Cantabria. Avenida de Los Castros s/n. E-39005 Santander, Spain
[3] School of Physics. University of New South Wales. Kensington NSW 2033, Australia

**Abstract.** We present the results of our study on three close pairs of QSOs. Our results are consistent with the existence of a proximity effect due to the foreground QSO, but due to its weakness we can only reject the absence of such effect at $\sim 1\sigma$ level. By modelling this proximity effect in terms of a simple photoionisation model, we find the best value for the UV ionizing background to be $\sim 10^{-20.5}$ erg cm$^{-2}$ s$^{-1}$ Hz$^{-1}$ srad$^{-1}$ at the Lyman limit, and an absolute lower limit (95% confidence) of $10^{-21.8}$ erg cm$^{-2}$ s$^{-1}$ Hz$^{-1}$ srad$^{-1}$. This lower limit rejects a number of models for the UV background where it is mostly contributed by QSOs and absorption by Lyman limit systems is taken into account.

## 1 Introduction

The inverse effect on the QSO Lyman $\alpha$ forest is a well known feature. Several models have been suggested to account for this effect. Some of them point to some kind of local mechanism around the observed QSO (e.g. gravitational infall or photoionisation), while other ones postulate conditions that are totally extrinsic to the QSO itself, as could be the case for small absorbing clouds –unable to totally cover the QSO continuum emitting region when located close to it– or a flattening of the distribution of lines at high redshift. Studying the existence of the inverse effect caused by a QSO close to the line of sight towards a background one in the spectrum of the latter will allow us the distinction between these two families of models.

**Table 1.** Observational data

| OBJECT | COORDINATES (J2000) | | | | | | $\theta(')$ | $z$ | $m_V$ | WAVELENGTH RANGE (Å) |
|---|---|---|---|---|---|---|---|---|---|---|
| | RA ($\alpha$) | | | Decl. ($\delta$) | | | | | | |
| 1055+021 | 10 | 57 | 57 | 01 | 54 | 03 | | 2.73 | 17.8 | |
| 1055+021 | 10 | 57 | 13 | 01 | 47 | 55 | 12.4 | 2.29 | 20.0 | 3910-4111 |
| 1222+228 | 12 | 25 | 27 | 22 | 35 | 13 | | 2.051 | 15.5 | |
| 1222+228 | 12 | 25 | 24 | 22 | 31 | 28 | 3.8 | 1.87 | 19.0 | 3255-3695 |
| 1228+077 | 12 | 31 | 21 | 07 | 25 | 18 | | 2.391 | 17.6 | |
| 1228+077 | 12 | 31 | 08 | 07 | 24 | 39 | 3.3 | 1.878 | 17.5 | 3360-3845 |

Three of such pairs have been observed (see Tab. 1), and line lists –with parameters $N$, $b$ and $z$ for each line– have been obtained by absorption line fitting. We also performed correlation analysis to eliminate lines belonging to metal systems.

## 2    Results

Following the suggestion by Bajtlik et al. (1988) we have modelled the proximity effect in terms of the enhancement of the UV photoionizing flux ($J_\nu$) by the nearby QSO. Thus, the model will allow us to estimate a value for the ratio of QSO UV flux to background flux –local to the cloud.

The model predicts that the distribution of Lyman $\alpha$ lines in the background QSO will have a 'hole' in the neighbourhood of the foreground one. The relative importance of this clearing will depend –as the *only* free parameter– on $J_\nu$. We can fit this function to our data using Maximum Likelihood methods to obtain the best value for $J_\nu$.

As this method turns out to produce a non-normalizable likelihood function for the values of $J_\nu$, we have also used other statistical tools (mainly Bootstrap and Monte Carlo methods). All of them have led us to the same final values for $J_\nu$.

This procedure finally allows us to marginally detect the Proximity Effect in our sample, but unfortunately the effect is so weak that we cannot reject its absence – there is a probability $\simeq 35\%$ of obtaining a result compatible with ours by chance, without any photoionization effect.

The best value that we obtain with our approach for the UV ionizing background is $\sim 10^{-20.5}$erg cm$^{-2}$ s$^{-1}$ Hz$^{-1}$ srad$^{-1}$ at the Lyman limit, and there is an absolute lower limit of $10^{-21.8}$erg cm$^{-2}$ s$^{-1}$ Hz$^{-1}$ srad$^{-1}$ (95% confidence). These values are compatible with other determinations of $J_\nu$ based on the inverse effect at similar redshifts (Bechtold et al. 1987, Bechtold 1995 – this volume–), and we agree with those works in pointing that this flux is stronger than it should be in order to explain all of it with the present models for QSOs (Bechtold et al. 1987, Madau 1992, Meiksin & Madau 1993, Madau 1995 –this volume–).

More details on this work can be found in Fernández-Soto et al. (1994).

## References

Bajtlik, S., Duncan, R.C., Ostriker, J.P. (1988): ApJ, **327**, 570
Bechtold, J., Weymann, R.J., Lin, Z., Malkon, M.A. (1987): ApJ, **315**, 180
Fernández-Soto, A., Barcons, X., Carballo, R., Webb, J.K. (1995): MNRAS, submitted
Madau, P. (1992): ApJ, **389**, L1
Meiksin, A. & Madau, P. (1993): ApJ, **412**, 34